\documentclass{nature_preprint}

\usepackage{graphicx}

\usepackage{caption}
\usepackage{times}
\usepackage{amsmath}
\usepackage{physics}
\usepackage{color}
\usepackage{upgreek}

\title{Controlled spatial separation of spins and coherent \\ dynamics in spin-orbit-coupled nanostructures}

\author{Shun-Tsung Lo$^{1\dagger}$, Chin-Hung Chen$^{1\dagger}$, Ju-Chun Fan$^{1\dagger}$, L. W. Smith$^{2\ddagger}$, G. L. Creeth$^{3}$, Che-Wei Chang$^{1}$, M. Pepper,$^{3}$ J. P. Griffiths$^{2}$, I. Farrer$^{2\S}$, H. E. Beere$^{2}$, G. A. C. Jones$^{2}$, D. A. Ritchie$^{2}$, \& T.-M. Chen$^{1\ast}$}

\begin{document}

\maketitle

\begin{affiliations}
 \item Department of Physics, National Cheng Kung University, Tainan 701, Taiwan
 \item Cavendish Laboratory, J J Thomson Avenue, Cambridge CB3 0HE, United Kingdom
 \item Department of Electronic and Electrical Engineering, University College London, London WC1E 7JE, United Kingdom
\end{affiliations}
\noindent $^\dagger$These authors contributed equally to this work.\\
\noindent $^\ddagger$Present Address: Wisconsin Institute for Quantum Information, University of Wisconsin-Madison, Madison, WI 53706, USA\\
\noindent $^\S$Present Address: Department of Electronic \& Electrical Engineering, University of Sheffield, Mappin Street, Sheffield S1 3JD, United Kingdom\\
\noindent $^\ast$To whom correspondence should be addressed; E-mail:  tmchen@mail.ncku.edu.tw.\\

\begin{abstract}

The spatial separation of electron spins followed by the control of their individual spin dynamics has recently emerged as an essential ingredient in many proposals for spin-based technologies because it would enable both of the two spin species to be simultaneously utilized, distinct from most of the current spintronic studies and technologies wherein only one spin species could be handled at a time. Here we demonstrate that the spatial spin splitting of a coherent beam of electrons can be achieved and controlled using the interplay between an external magnetic field and Rashba spin-orbit interaction in semiconductor nanostructures. The technique of transverse magnetic focusing is used to detect this spin separation. More notably, our ability to engineer the spin-orbit interactions enables us to simultaneously manipulate and probe the coherent spin dynamics of both spin species and hence their correlation, which could open a route towards spintronics and spin-based quantum information processing.

\end{abstract}

The spin-orbit interaction in materials gives rise to a separation of different spin species in momentum space, creating many interesting phenomena such as the spin Hall\cite{Hirsch1999,Kato2004,Valenzuela2006}, the quantum spin Hall\cite{Bernevig2006a,Bernevig2006} effects, and the spin-momentum locking\cite{Hsieh2009,Li2014}. However, it does not separate the spin up and spin down electrons in real space. In other words, even though different spins behave very differently they cannot be resolved and tracked in real space, similar to spin-degenerate systems where the spin-orbit interaction is negligible. So far most of the spintronic technologies which require spin to be resolved prior to subsequent operations have to rely on the creation of a spin imbalance with, for example, ferromagnets or optical injection. However, these methods are limited in both fundamental and practical aspects since only one spin type (i.e., the majority spin) can be utilized. For example, the correlation between different spin types remains experimentally unexplored unless one can resolve and track both spin types simultaneously, for which it is necessary to spatially split electron spins rather than polarize them. Developing a simple way to spatially separate the opposite spin types, then manipulate and track the coherent spin dynamics of both of the two spin types and, more importantly, their phase correlation is therefore essential and a frontier in current research.

The Stern-Gerlach magnet is well-known for separating spins but is limited to uncharged particles, and modified proposals for electron spins using inhomogeneous spin-orbit effective fields\cite{Ionicioiu2003,Ohe2005,Kohda2012} have yet to be realized. The spin Hall effect geometry\cite{Hirsch1999,Kato2004,Valenzuela2006} can also produce spin separation, where the diffusive electrons that are scattered to opposite edges of a conductor are coupled to spins of opposite orientations; however, no control can be exercised in such a random scattering system. A promising way to achieve spatial separation of electron spins in a spin-orbit coupled system is to apply a transverse magnetic field. Spin-up and spin-down electrons have different momenta and thus, when moving through a magnetic field, will experience different Lorentz forces and consequently undergo different cyclotron motions. This concept has been successfully demonstrated, using a hole gas in which the spin-orbit interaction was not tunable\cite{Rokhinson2004,Rokhinson2006,Chesi11}, but to manipulate and study the behaviour of the spatially separated spins remains an outstanding challenge.

Here we combine this simple concept of spatial spin separation with techniques to coherently manipulate and detect spins, and thereby demonstrate a spatial spin splitting of a coherent electron beam together with full control of the dynamics of these spatially separated spins. The spatial separation, coherent spin dynamics, and phase correlation between the up- and down-spin electrons can all be -- electrically and on-chip -- controlled and probed. This allows both of two spin types (instead of just the majority one as in most previous studies) to be simultaneously probed and manipulated, which promises to advance spintronic technologies that require both spin types to be operated together.

\section*{Results}
\subsection{Spatial separation of spins} Figure~1a captures the operation of our devices. A quantum point contact (QPC) -- a one-dimensional constriction created by applying voltages to split gates patterned on the surface of an InGaAs heterostructure -- is used to inject an unpolarized electron beam into a two-dimensional electron gas (2DEG). The 2DEG is formed in the InGaAs quantum well (see Methods), wherein the structural inversion asymmetry of the well generates a momentum-dependent magnetic field $\textbf{B}^{\text{SO}}_{\text{R}}$ on the spin of every moving electron, the so-called Rashba spin-orbit interaction. This Rashba spin-orbit effective magnetic field $\textbf{B}^{\text{SO}}_{\text{R}}$ lies in the plane of the 2DEG (i.e., the $x$-$y$ plane in Fig.~1a) and is orientated perpendicular to the electron's momentum. It lifts the spin degeneracy in momentum space and leads to two spin-polarized Fermi circles, parallel and antiparallel to $\textbf{B}^{\text{SO}}_{\text{R}}$ (Fig.~1b). Electrons in the parallel and antiparallel spin states (hereafter, we refer to these as the up and down spins, respectively), though moving in the same direction and spatially unresolved when injected from a QPC into the 2DEG, have different Fermi wavevectors and thus will be deflected along different cyclotron trajectories in the presence of a transverse magnetic field. Spin-selective spatial separation of an electron beam is therefore achieved.

To study the spatial separation of the two spin species, another QPC is placed at a distance $L$ from the QPC emitter to act as a charge collector, forming a geometry (Figs.~1a, 1c, and the inset of 1d) known as transverse magnetic focusing\cite{Tsoi74,Folk03,Rokhinson2004,Rokhinson2006,Aidala2007,Chen2012,Taychatanapat2013}. Magnetic focusing occurs when the electrons that leave the QPC emitter are focused into the QPC collector, giving peaks in collector voltage (i.e., focusing peaks) at magnetic fields where an integer multiple of cyclotron diameter is equal to $L$. The two spatially separated spin species travel with different cyclotron radii and thus will require two different magnetic fields
\vspace*{3mm}\begin{equation}
B_{\uparrow \downarrow} = \frac{2\hbar k_{\uparrow \downarrow}}{eL} = \frac{2(\sqrt{2m^* E_{\text{F}}}\mp m^* \alpha / \hbar)}{eL},
\label{eq1}
\vspace*{3mm}\end{equation} 
to focus themselves directly into the collector (inset of Fig.~1d), where $\hbar$ is Planck's constant divided by $2\pi$, $e$ is the elementary charge, $m^*$ is the electron effective mass, $E_{\text{F}}$ is the Fermi energy, $k_{\uparrow}$ ($k_{\downarrow}$) refers to the Fermi wavevector of spin-up (-down) state, and $\alpha$ parametrizes the strength of Rashba spin-orbit interaction. A spatial splitting of electron spins is therefore visible as a peak splitting in the magnetic focusing spectrum, allowing us to easily track and investigate the spatial spin separation.

Figure~1d shows the magnetic focusing spectrum, with the emitter ($G_{\text{E}}$) and collector conductance ($G_{\text{C}}$) both set to $100$~$\upmu$S (above the quantized plateau at $2e^2/h$) to allow both spin species to propagate through the one-dimensional channels (Supplementary Notes~1 and 2). For $B < 0$ focusing peaks appear periodically at integer multiples of $B \approx 0.19$~T, corresponding to when electrons are focused into the collector. This value is consistent with the cyclotron motion $B = 2\sqrt{2m^* E_{\text{F}}}/eL$ calculated using the two-dimensional electron density. For $B > 0$ electrons are directed in the opposite direction, therefore no peaks in collector voltage are observed. The splitting of the focusing peak (hereafter referred to as the focusing peak doublet) is observed on the first and the third focusing peaks as evidence of spatial spin splitting. The low-field $B_\uparrow$ and high-field $B_\downarrow$ peak within the doublet corresponds to the spin-up and spin-down electrons, respectively. The Rashba parameter $\alpha$ estimated from the peak splitting using equation~(1) is $3.1 \times 10^{-11}$~eVm, close to the value estimated from the beating pattern in the Shubnikov-de Haas oscillations (see Supplementary Note~1). There is additional structure around the focusing peaks which is likely due to the quantum interference effects\cite{VanHouten1989}. We note that the focusing peak doublet is not visible on the second focusing peak. This is consistent with the model\cite{Usaj2004} that the electrons are subject to spin-flip when they are reflected from the edge of the 2DEG and hence the two spatially separated spin branches reunite with each other at the collector (see Supplementary Note~3).

\subsection{Control of charge and spin dynamics} So far the magnetic focusing spectrum can only show that the electrons leaving from a QPC emitter are spatially spin split, without being able to shed any light on the spin dynamics afterwards. An important open question remains on how each spin species evolves due to the influence of a rotating $\textbf{B}^{\text{SO}}_{\text{R}}$ (in the reference frame of the spin) -- which rotates along the cyclotron trajectories as the momentum rotates -- in such a spin-orbit coupled 2DEG. For example, it is desirable to understand whether electron spins can maintain their coherence before reaching the collector, and also whether these spins adiabatically follow $\textbf{B}^{\text{SO}}_{\text{R}}$. To study the binary spin dynamics, we now force the collector to act as a spin analyzer by introducing the lateral spin-orbit interaction\cite{DebrayP2009,Chuang2015} and manipulating the energy and population of the one-dimensional subbands in the collector. A voltage difference between the two sides of the split gate is used to create a lateral inversion asymmetry and consequently a lateral spin-orbit effective magnetic field $\textbf{B}^{\text{SO}}_{\text{L}}$ pointing along the $z$ axis (Fig.~1a). The electrically tunable $\textbf{B}^{\text{SO}}_{\text{L}} + \textbf{B}^{\text{SO}}_{\text{R}}$ within the emitter and collector QPC allows us to respectively prepare and analyze the electron spins along any specific direction in the $y$-$z$ plane. Additionally, a top gate (gate T in Fig.~1c) covers the entire focusing path and is used to vary $\textbf{B}^{\text{SO}}_{\text{R}}$ (and equivalently $\alpha$) in the 2DEG region.

The electron spins transmitted through the QPC emitter stabilize at the state determined by $\textbf{B}^{\text{SO}}_{\text{L}} + \textbf{B}^{\text{SO}}_{\text{R}}$ and consequently their orientations are initialized out of the 2DEG plane. In other words, the spin-up (spin-down) electrons are tilted toward negative (positive) $z$-direction by $\textbf{B}^{\text{SO}}_{\text{L}}$ owing to being in the one-dimensional $\textbf{B}^{\text{SO}}_{\text{L}} + \textbf{B}^{\text{SO}}_{\text{R}}$ parallel (antiparallel) spin states. After leaving the emitter, the electrons experience only the in-plane $\textbf{B}^{\text{SO}}_{\text{R}}$ (since the focusing transverse magnetic field is small compared to $\textbf{B}^{\text{SO}}_{\text{R}}$) and therefore can precess about it as depicted in Fig.~1a if they propagate coherently. We can alter the spin orientation by controlling the spin precession frequency using top gate voltage $V_\text{T}$, which determines $\textbf{B}^{\text{SO}}_{\text{R}}$. Here we first demonstrate an electrically-tunable spatial spin separation in Fig.~2a, where the evolution of focusing spectrum of the first doublet is measured as a function of $V_{\text{T}}$ at $G_{\text{E}} = 160$~$\upmu$S and $G_{\text{C}} = 100$~$\upmu$S. The two superimposed dashed lines are the calculated $B_\uparrow$ and $B_\downarrow$ focusing fields using the model of spin precession described below. The spatial separation between the two spin species, manifested as the peak splitting $|B_{\downarrow} - B_{\uparrow}| = 4m^* \alpha / \hbar eL$, increases with increasing $V_{\text{T}}$.

We now move on to study the spin dynamics of the two spin species and the phase correlation between them. This is achieved by lowering the collector conductance to $G_{\text{C}} = 20$~$\upmu$S such that the QPC acts as a spin analyzer, as described in Supplementary Note~2. The orientation of incident electron spins is indicated by the magnitude of the collector voltage (i.e., the focusing peak height). Electrons can propagate through the collector if their spin is parallel to the polarization direction, and cannot pass if their spin is antiparallel. Figure~2b shows that both the $B_\uparrow$ and $B_\downarrow$ focusing peaks in collector voltage oscillate with $V_{\text{T}}$. These oscillations are $\pi$ out-of-phase with each other, i.e., each local maximum (minimum) in collector voltage along the $B_\uparrow$ focusing peak -- which corresponds to the incident spins being parallel (antiparallel) to the polarization direction of the collector -- coincides with the local minimum (maximum) along the $B_\downarrow$ peak. Evidently, both the up and down spin coherently precess and maintain their initial $\pi$ out-of-phase correlation after undergoing the action of the rotating $\textbf{B}^{\text{SO}}_{\text{R}}$.

Within the adiabatic approximation in which $\textbf{B}^{\text{SO}}_{\text{R}}$ changes its direction slowly such that the system adapts its configuration accordingly, the direction of electron spins with respect to $\textbf{B}^{\text{SO}}_{\text{R}}$ remains conserved (i.e., the spinors can be described as a superposition of the adiabatic $\textbf{B}^{\text{SO}}_{\text{R}}$ eigenstates with conserved probabilities; see Supplementary Note~4 for more details). Hence, the electron spins precess about $\textbf{B}^{\text{SO}}_{\text{R}}$ with a Larmor frequency of $\omega^s_{\uparrow \downarrow} = 2 \alpha |k_{\uparrow \downarrow}|/\hbar$. The spin precessional angle accumulated by electrons traveling along a semiclassical cyclotron orbit to the collector is therefore given as\cite{Datta1990} 
\vspace*{3mm}\begin{equation}
\phi (V_{\text{T}}) = \omega^s_{\uparrow \downarrow} t_{\uparrow \downarrow} = \pi m^* \alpha (V_{\text{T}}) L/ \hbar^2,
\label{eq2}
\vspace*{3mm}\end{equation}
where $t_{\uparrow \downarrow}$ is the time interval for the focusing process. This angle is irrespective of spin orientation and depends only on the strength of spin-orbit interaction $\alpha(V_\text{T})$ for a fixed $L$, consistent with the observation of antiphase oscillations in the collector voltage for $B_\uparrow$ and $B_\downarrow$ in Fig.~2b. Moreover, the oscillations enable us to calculate the gate-voltage dependent variation of $\alpha$ using equation~(2), which is consistent with the value obtained from the splitting of focusing peaks using equation~(1). Later in this paper we will compare these $\alpha$ values derived independently using these two methods. Such a consistency is here evident from the excellent quantitative agreement between the position of the focusing peaks measured experimentally and the values calculated using equation~(1) in accordance with the $\alpha$ value derived from the spin precessional motion (dashed lines in Figs.~2a and 2b). The fact that the two antiphase oscillations are quantitatively described by considering spin precession in the adiabatic limit indicates that both of the spatially separated up and down spin adiabatically follow and precess about the rotating $\textbf{B}^{\text{SO}}_{\text{R}}$ as illustrated in Fig.~1a. This also suggests that the phase correlation between the two separate, neighboring spin types can be electrically controlled via tuning $\alpha$. It is worth noting that the phase correlation observed in focusing spectra is equal to $\pi$ regardless of the strength of spin-orbit interaction because both spin types are focused into the same collector by different magnetic fields, while in reality opposite spins travel along different trajectories and gain different phase shifts determined by equation~(2).

Similar results are observed in other devices, as shown in Fig.~3 where the data are obtained using device~B after illumination (see Methods). Figures~3a and 3b compare the magnitude of the $B_\uparrow$ and $B_\downarrow$ focusing peaks as a function of $V_{\text{T}}$, for $G_{\text{C}}=20$ and $100$~$\upmu$S, respectively. For $G_{\text{C}}=20$~$\upmu$S (Fig.~3a) the collector acts as a spin analyzer. The $B_\uparrow$ and $B_\downarrow$ collector voltages oscillate with $V_{\text{T}}$, and are $\pi$ out of phase with each other. In contrast, no oscillations are observed when $G_{\text{C}}$ is raised to $100$~$\upmu$S, where the QPC collector acts only as a charge detector (Fig.~3b). The oscillations also disappear when either the emitter or the collector QPC is biased symmetrically (Supplementary Note~5), which is consistent with our spin precession model. When the emitter is biased symmetrically (i.e., $\textbf{B}^{\text{SO}}_{\text{L}} = 0$), the electron spins which are emitted are aligned along the axis of $\textbf{B}^{\text{SO}}_{\text{R}}$ and hence no spin precession shall occur. Also, when $\textbf{B}^{\text{SO}}_{\text{L}}$ is removed from the collector, the spin polarization is analyzed along the stationary $\textbf{B}^{\text{SO}}_{\text{R}}$ spin states, and hence no spin precession can be probed. Note that as with device~A, there is a quantitative agreement between the $\alpha (V_{\text{T}})$ obtained with equations~(1) and (2), which use the peak splitting and oscillatory collector voltage data, respectively.

One advantage of device~B is that the QPC emitter and collector can be independently controlled since they do not share a common middle gate. This enables us to reverse the polarity of the lateral inversion asymmetry of the QPC and hence $\textbf{B}^{\text{SO}}_{\text{L}}$, simply by reversing the polarity of the voltage difference between the two sides of the split gate. Figure~3c,~d presents a comparison of the focusing spectra for $\Delta V_\text{E} = +1.5$~V and $-1$~V. Here the focusing spectra are plotted as a function of magnetic field and $\alpha(V_{\text{T}}) \times L$, instead of $V_{\text{T}}$ (as in other figures), since when $\Delta V_\text{E}$ changes the distance $L$ between the emitter and collector also changes and thus needs to be taken into account (see Supplementary Note~6). A phase inversion in the oscillations for both the $B_{\uparrow}$ and $B_{\downarrow}$ focusing peaks is apparent as the lateral bias $\Delta V_\text{E}$ is changed from $+1.5$~V to $-1$~V. Such an inversion can be easily understood using a schematic in Fig.~3e which illustrates the phase evolution, depicted using Bloch spheres, of the spin-up (red arrows) and spin-down (blue arrows) electrons travelling along the cyclotron trajectory at positive and negative $\Delta V_{\text{E}}$, respectively. Since the initial phase correlation between spin-up and spin-down electrons is inverted as the direction of $\textbf{B}^{\text{SO}}_{\text{L}}$ is reversed, the observed phase correlation for the arrivals that undergo the same phase evolution (and equivalently $\alpha L$) must also be inverted.

Figure 4 summarizes values obtained for the Rashba coefficient $\alpha$. The values obtained via the focusing peak splitting using equation~(1) (open symbols) and via the oscillatory collector voltage using equation~(2) (solid symbols) are both shown and are in excellent quantitative agreement with each other. In order to directly compare data before (red symbols) and after (blue symbols) illumination, we plot the Rashba coefficient $\alpha$ as a function of carrier density. The value of $\alpha$ follows the same trend line both before and after illumination. For comparison, we also plot the values of $\alpha(V_{\text{T}})$ published in recent work\cite{Chuang2015} using a spin field-effect transistor (fabricated on the same wafer used here), where $\alpha$ is estimated from spin precession measurements in a steady -instead of rotating- $\textbf{B}^{\text{SO}}_{\text{R}}$. There is excellent quantitative agreement between the values of $\alpha$ obtained from these different devices and methods. The spin focusing technique appears more informative than the conventional SdH beating analysis\cite{Nitta1997} which is sometimes difficult to observe (Supplementary Note~1), and provides a reliable means for the determination of $\alpha$ value in the ballistic transport regime.

\section*{Discussion}
The ability to manipulate and probe coherent spin dynamics in materials with high spin-orbit interaction is important for understanding the physics of emerging materials, as well as to having implications for spintronics and (topological) quantum computing\cite{Dario2015,Petersson2012,Nayak2008}. Distinct from most previous studies\cite{Chuang2015,Koo2015} -- which rely on the introduction of polarized electrons to break the spin symmetry and are limited in that only the majority spin type can be resolved and used -- our spin focusing technique provides a route to probe and manipulate the coherent spin dynamics of both spin species and their phase correlation in semiconductor nanostructures, and can be readily extended to materials with unusual band structures such as topological insulators\cite{Hsieh2009,Li2014,Hasan2010}, graphene and its hybrid structures\cite{Geim2013}. A recent study\cite{Taychatanapat2013} that used the conventional magnetic focusing technique to probe the properties of graphene is a successful example. From a technological viewpoint, our ability to spatially bifurcate the two electron spin types and coherently manipulate them to any specific orientation (through spin precession and the fast manipulation of $\textbf{B}^{\text{SO}}_{\text{R}}$ and $\textbf{B}^{\text{SO}}_{\text{L}}$ using surface gates) make it possible to prepare two separate, neighbouring spins with an electrically controllable phase correlation, which has implications for interferometer and quantum logic operations.

\section*{Methods}
\subsection{Devices.} A gated modulation-doped In$_{0.75}$Ga$_{0.25}$As/In$_{0.75}$Al$_{0.25}$As heterostructure is used in this work. The layer sequence is grown by molecular beam epitaxy as follows: 250 nm In$_{0.75}$Al$_{0.25}$As; 30 nm In$_{0.75}$Ga$_{0.25}$As (quantum well); 60 nm In$_{0.75}$Al$_{0.25}$As (spacer); 15 nm In$_{0.75}$Al$_{0.25}$As (Si-doped); 45 nm In$_{0.75}$Al$_{0.25}$As; and 2 nm In$_{0.75}$Ga$_{0.25}$As (cap). A dielectric layer (27 nm and 40 nm for device~A and B, respectively) of SiO$_2$ is deposited on the wafer surface by plasma-enhanced chemical vapor deposition. Subsequently, surface gates are defined using electron-beam lithography and thermal evaporation of Ti/Au. There are two device designs, denoted device~A and device~B, as shown in Fig.~1c. In device~A the lateral biases of the emitter and collector QPCs are defined as $\Delta V_\text{E} = V_\text{E}-V_\text{M}$ and $\Delta V_\text{C} = V_\text{C}-V_\text{M}$, respectively, whereas in device~B $\Delta V_\text{E} = V_\text{E1}-V_\text{E2}$ and $\Delta V_\text{C} = V_\text{C1}-V_\text{C2}$. Note that the emitter is covered by the top gate, such that the Fermi wavevector of the focusing electrons that transit from the emitter to the bulk can be reliably controlled with the top gate. The collector is not covered by the top gate so that the spin polarization can be analyzed along a fixed axis, independent of the top gate voltage. Data from device~A and B are taken before and after illumination, respectively, which give very different characteristics of the 2DEG.

\subsection{Measurements.} Experiments are performed at a base temperature of 25 mK in a dilution refrigerator equipped with a superconducting magnet. The carrier density and mobility of the 2DEG are measured to be $2.1\times10^{11}$~cm$^{-2}$ and $1.7\times10^5$~cm$^2$V$^{-1}$s$^{-1}$, respectively, using four-terminal magnetotransport measurements (Supplementary Note~1). This gives a mean free path of $1.3$~$\upmu$m for momentum relaxation. After illumination, they increased to $3.9\times10^{11}$~cm$^{-2}$, $2.6\times10^5$~cm$^2$V$^{-1}$s$^{-1}$, and $2.7$~$\upmu$m, respectively. For transverse magnetic focusing experiments, simultaneous lock-in measurements of emitter and collector QPC conductances are carried out by supplying two-independent excitation sources of a $77$~Hz a.c. voltage $V_{\text{exc}} = 100$~$\upmu$V to the emitter and a $37$~Hz a.c. current $I_{\text{exc}} = 1$ nA to the collector. The magnetic field is applied normal to the 2DEG plane to focus electrons into the collector. The focusing signal is measured as a voltage drop developed across the QPC collector in linear response to the 77 Hz a.c. current from the QPC emitter.

\subsection{Data availability.} The data that support the findings of this study are available from the corresponding author upon reasonable request.


\section*{Acknowledgements}
We thank S.-C. Ho and C.-T. Liang for discussions. This work was supported by the Ministry of Science and Technology (Taiwan), the Headquarters of University Advancement at the National Cheng Kung University, and the Engineering and Physical Sciences Research Council (UK).

\section*{Author Contributions}
S.-T.L., C.-H.C., and J.-C.F. performed the measurements and analysed the data, in which C.-W.C. and T.-M.C. participated. L.W.S. and G.L.C. fabricated the devices with contributions from M.P., and T.-M.C; I.F., H.E.B., and D.A.R. provided wafers; J.P.G. and G.A.C.J. performed electron-beam lithography. S.-T.L and T.-M.C. wrote the paper with input from all of the authors; T.-M.C. designed and coordinated the project.

\section*{Additional information}
\textbf{Competing interests:} The authors declare no competing financial interests.

\newpage

\begin{figure}
\begin{center}
\includegraphics[width=0.95\columnwidth]{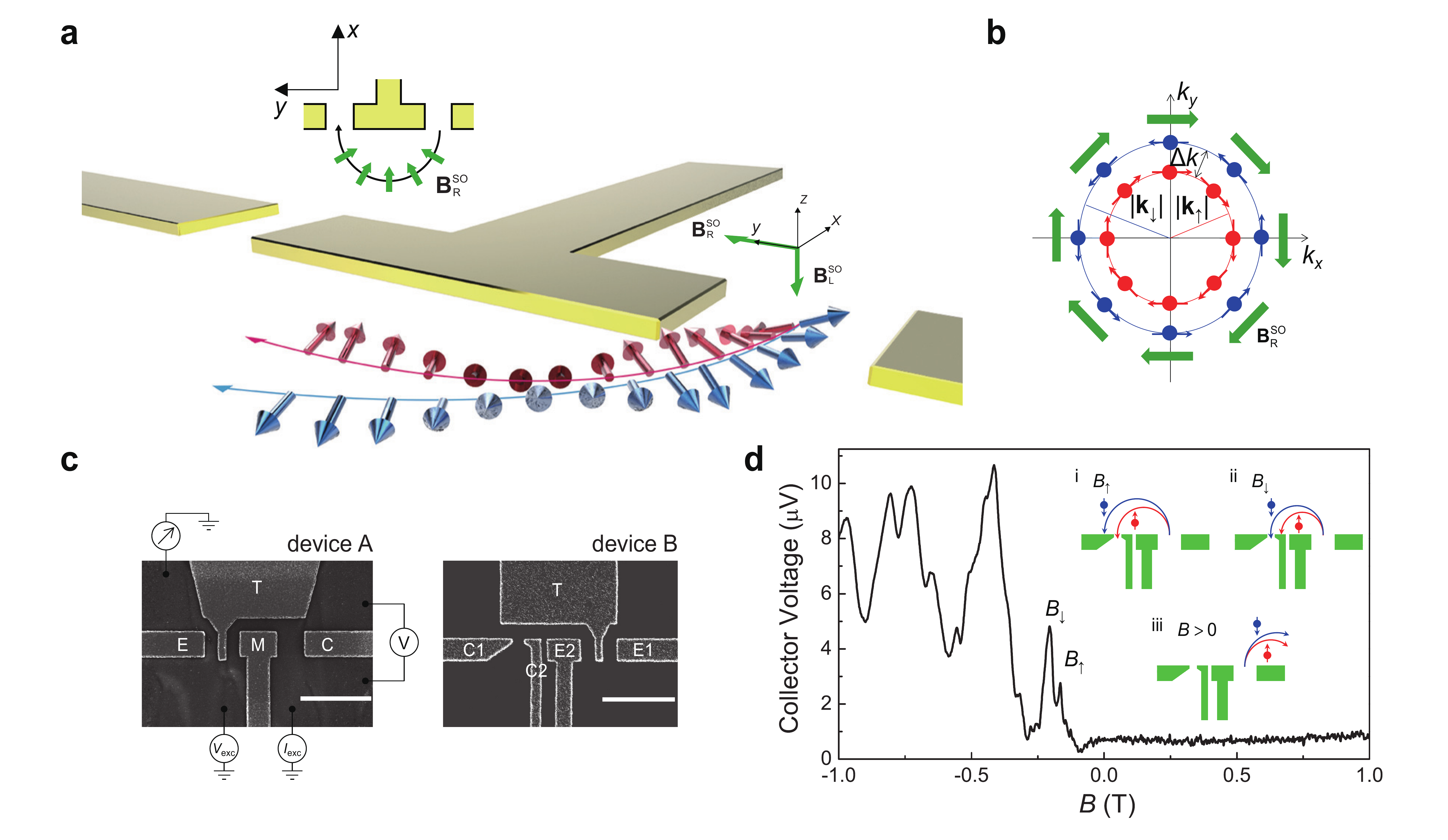}
\end{center}
\noindent {Figure~1:} \textbf{Scheme for spatial spin separation and control of spin dynamics.} \textbf{a}, Schematic view of a spin focusing device. The structural inversion asymmetry gives rise to an in-plane Rashba spin-orbit field $\textbf{B}^{\text{SO}}_{\text{R}}$ on the spin of every moving electron, illustrated by the inset. We define the spin up, $\uparrow$ (spin-down, $\downarrow$), as parallel (antiparallel) to $\textbf{B}^{\text{SO}}_{\text{R}}$. Spin-up and spin-down electrons have different Fermi wavevectors and thus will be deflected along different cyclotron trajectories in a transverse magnetic field, resulting in spatial spin separation. Within the QPC constriction, an additional lateral spin-orbit field $\textbf{B}^{\text{SO}}_{\text{L}}$ can be created via laterally biasing the gates to tilt spins toward either positive or negative $z$ direction. The two spatially separated spin species thus precess about $\textbf{B}^{\text{SO}}_{\text{R}}$ in the 2DEG region. The spin-orbit fields $\textbf{B}^{\text{SO}}_{\text{R}}$ and $\textbf{B}^{\text{SO}}_{\text{L}}$ are represented by green arrows, while the red and blue arrows represent up and down spins, respectively. \textbf{b}, The Fermi surface (red and blue circle of radius $k_{\uparrow}$ and $k_{\downarrow}$ for spin up and spin down) is spin-split with a wavevector separation $\Delta k$ ($= k_{\downarrow} - k_{\uparrow}$) in the presence of Rashba spin-orbit interaction. The arrows are coloured following the same convention as in \textbf{a}. \textbf{c}, SEM images of device~A and B, with scale bar of 1 $\upmu$m. Devices~A and B are measured before and after illumination, respectively, which gives markedly different electron densities and mobilities (see Methods). Device~B contains two pairs of split gates to allow independent control of the QPC emitter (using E1 and E2) and collector (using C1 and C2). \textbf{d}, Transverse magnetic focusing spectrum measured from device~B. The inset shows representative trajectories for spin-up (red trace) and spin-down (blue trace) electrons at different magnetic fields.
\end{figure}

\begin{figure}
\begin{center}
\includegraphics[width=0.9\columnwidth]{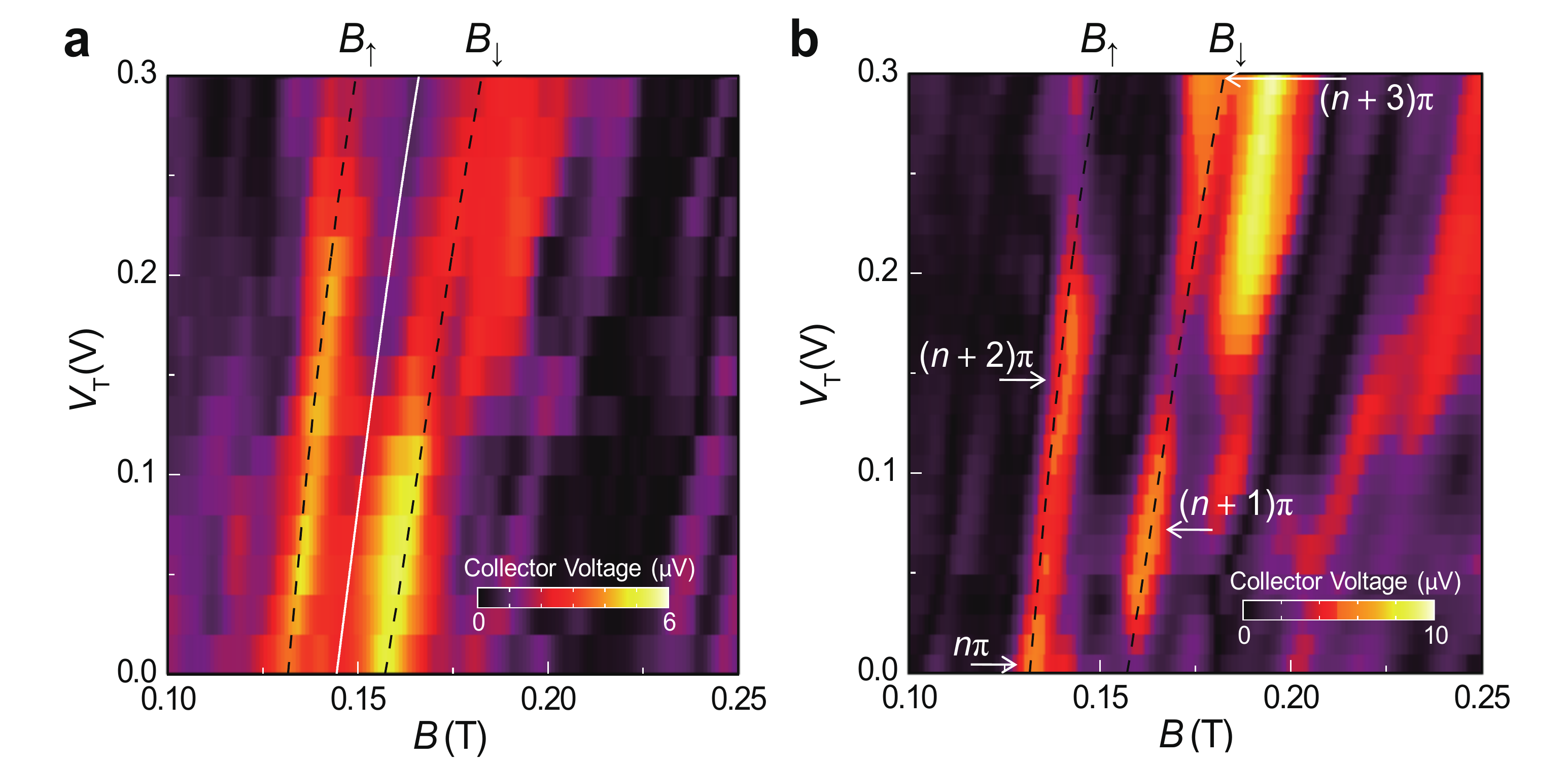}
\end{center}
\noindent {Figure~2:} \textbf{Magnetic spin focusing spectra.} \textbf{a}, Collector voltage as a function of magnetic field $B$ and top gate voltage $V_{\text{T}}$ for device~A with emitter conductance $G_{\text{E}} = 160$~$\upmu$S and collector conductance $G_{\text{C}} = 100$~$\upmu$S. The lateral bias $\Delta V_\text{E}$ is fixed at $1.33$~V (see Methods for the quantification of $\Delta V_\text{E}$) whereas $\Delta V_\text{C}$ ranges from $2.15$~V to $2.41$~V as $V_{\text{T}}$ increases to keep both QPCs at fixed conductance values. The solid line illustrates the average $B$ between the spin-up and spin-down focusing peaks $(B_{\uparrow} + B_{\downarrow})/2$, which can be used to determine the carrier density $n_{\text{2D}}$. The dashed lines show the focusing peak positions calculated using the spin precessional motion. \textbf{b}, As in \textbf{a} but with $G_{\text{C}}$ reduced to $20$~$\upmu$S to turn the collector into a spin analyzer. $\Delta V_\text{E}$ is fixed at $1.23$~V whereas $\Delta V_\text{C}$ ranges from $2.02$~V to $2.30$~V. The subsequent maxima (minima) of the oscillating collector voltage along the $B_{\uparrow}$ and $B_{\downarrow}$ focusing peaks correspond to rotations of the incident spins by $n \pi$, where $n$ is an integer, such that the spin is parallel (antiparallel) to the polarization direction of the collector.
\end{figure}

\begin{figure}
\begin{center}
\includegraphics[width=1\columnwidth]{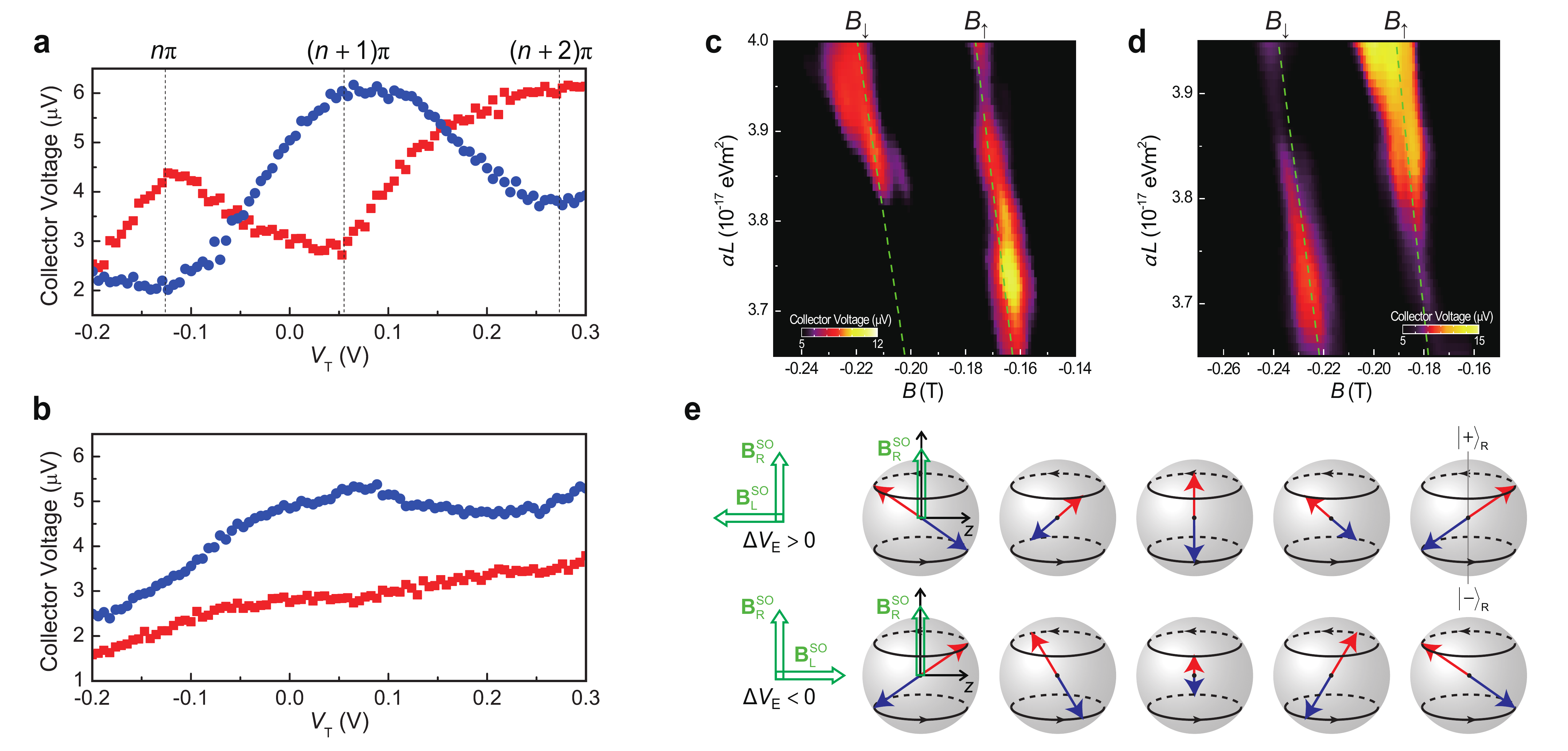}
\end{center}
\noindent {Figure~3:} \textbf{Spin precession in a rotating $\textbf{B}^{\text{SO}}_{\text{R}}$.} \textbf{a}, Collector voltage of the $B_{\uparrow}$ (red) and $B_{\downarrow}$ (blue) focusing peaks as a function of $V_{\text{T}}$, with $G_{\text{E}} = 100$~$\upmu$S and $G_{\text{C}} = 20$~$\upmu$S. Data for all panels in this figure are from device~B. The lateral biases of the QPC emitter and collector are set at $\Delta V_\text{E} = 0.25$~V and $\Delta V_\text{C} = 0.5$~V, respectively. \textbf{b}, As in \textbf{a} except with $G_{\text{C}}$ increased to $100$~$\upmu$S for comparison. \textbf{c}, Magnetic spin focusing spectrum as a function of $\alpha L$ and magnetic field for $\Delta V_\text{E} = 1.5$~V and $\Delta V_\text{C} = 0.5$~V. \textbf{d}, As in \textbf{c} but with $\Delta V_\text{E}$ changed to $-1$~V to invert the direction of $\textbf{B}^{\text{SO}}_{\text{L}}$. This gives rise to an inverse $\pi$ out-of-phase oscillation in the $B_{\uparrow}$ and $B_{\downarrow}$ focusing peaks with respect to that in panel \textbf{c}. Only the data with the collector voltage above 5~$\upmu$V are shown to highlight the varying focusing peak height. Data for panels \textbf{c} and \textbf{d} are obtained in a different cooldown to panels \textbf{a} and \textbf{b}. The dashed lines indicate the focusing peak positions calculated using the same method as in Fig.~2. \textbf{e}, A sequence of Bloch spheres illustrate the phase evolution of the spin-up (red arrows) and spin-down (blue arrows) electrons moving along the focusing trajectory. The top (bottom) row of spheres represents the phase evolution for $\Delta V_\text{E}>0$ ($\Delta V_\text{E}<0$); the vertical (horizontal) axis represents $\textbf{B}^{\text{SO}}_{\text{R}}$ ($\textbf{B}^{\text{SO}}_{\text{L}}$). Starting with electrons within the QPC emitter, the combination of Rashba and lateral spin-orbit interactions prepares the $\textbf{B}^{\text{SO}}_{\text{R}} + \textbf{B}^{\text{SO}}_{\text{L}}$ parallel and antiparallel spin states. After leaving the QPC and entering the 2DEG both spin types experience only the Rashba effective field $\textbf{B}^{\text{SO}}_{\text{R}}$ and precess about it.
\end{figure}

\begin{figure}
\begin{center}
\includegraphics[width=0.7\columnwidth]{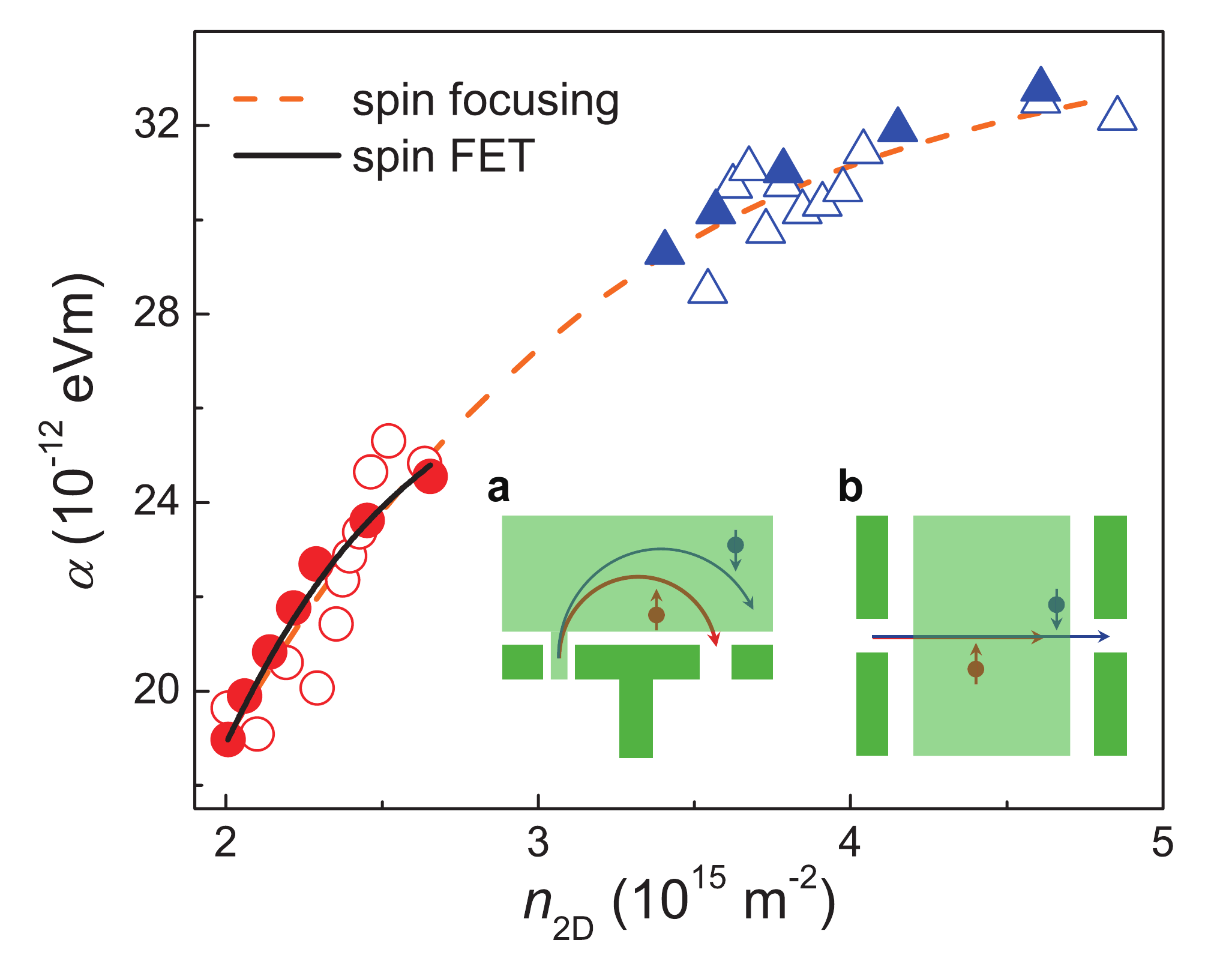}
\end{center}
\noindent {Figure~4:} \textbf{Comparison of the measured Rashba coefficients.} The Rashba coefficient $\alpha$ is plotted as a function of carrier density $n_{\text{2D}}$. Red and blue data points correspond to data obtained using the magnetic spin focusing technique (illustrated in inset \textbf{a}) before (device~A) and after (device~B) illumination, respectively. Two methods are used to extract $\alpha$. Open symbols show the values given by equation~(1) in the main article which considers the spatial spin separation of electrons. Solid symbols show values obtained using equation~(2) which considers the precessional motion of the spin. The dashed line shows a polynomial fit to the data from spin focusing. For comparison, the Rasha coefficient obtained in recent measurements of a spin field effect transistor\cite{Chuang2015} (illustrated in inset \textbf{b}) are shown by the black solid line.
\end{figure}

\end{document}